\documentclass{article}
\usepackage{ltwol2e}
\usepackage{amsmath}
\usepackage{amssymb} 
\usepackage{pst-node}
\usepackage{graphics}

\def\MeV{\mathrm{MeV}}
\def\GeV{\mathrm{GeV}}
\def\TeV{\mathrm{TeV}}
\def\pb{\mathrm{pb}}
\def\fb{\mathrm{fb}}
\def\ab{\mathrm{ab}}
\def\lumi{{\textstyle\int}\mathcal{L}}

\newcommand{\strt}{\rule[-5pt]{0pt}{14pt}}

\begin{document}

\title{ASPECTS OF HIGGS PHYSICS AND PHYSICS BEYOND THE STANDARD MODEL\\
AT LHC AND \boldmath$e^+e^-$ LINEAR COLLIDERS$^*$}

\author{W.\ KILIAN}
\address{Institut f\"ur Theoretische Physik, Universit\"at Heidelberg,
D--69120 Heidelberg, Germany\\E-mail: W.Kilian@thphys.uni-heidelberg.de}

\author{P.\ M.\ ZERWAS}
\address{Deutsches Elektronen-Synchrotron DESY,
D--22603 Hamburg, Germany\\E-mail: zerwas@desy.de}   


\twocolumn[\maketitle\abstracts{ %
Recent developments in prospects of searching for Higgs particles and
testing their properties at the LHC and at TeV $e^+e^-$ linear
colliders are summarized.  The discovery limits of supersymmetric
particles at the LHC are presented and the accuracy is explored with
which the fundamental SUSY parameters in the context of supergravity
models can be determined at high-luminosity linear colliders.
Finally, new discovery limits for gauge bosons in left-right symmetric
models at the LHC are presented.
}]

\footnotetext{$^*$ Summary of contributions to PA10 at ICHEP'98
(Vancouver).}

\section{The physical basis}

\textbf{1.}
Recent results from high-precision measurements of electroweak
observables at LEP, SLC and elsewhere strongly support the hypothesis
that the electroweak symmetries are broken through the Higgs-mechanism
and that a light fundamental Higgs boson is realized in Nature~\cite{Kar}.
Moreover, the perturbative expansion of the theory up to the GUT
scale, backed by the observed value of the electroweak mixing angle,
favors a Higgs mass  in the intermediate range $M_H\lesssim
180\;\GeV$.  This is the most difficult range to explore at
the LHC, yet it is evident now that the lower part of this
range can be covered in the $H\to\gamma\gamma$ decay
channel~\cite{Bar} while the upper part is accessible in the four-lepton
decay $H\to ZZ^*\to 4\ell^{\pm}$, both with high significance between
$S=8$ and~$10$.

Once Higgs bosons are discovered, their properties must be explored to
establish the Higgs mechanism \emph{sui generis} as the basic
mechanism for the breaking of the electroweak symmetries.  This
program can be carried out in three consecutive steps.  (i) The
external quantum numbers $J^{PC}$ must be determined~\cite{BFH}. (ii)
The generating of vector-boson and fermion masses through the Higgs mechanism
can be scrutinized by measuring the $HVV$ and $Hff$ couplings,
\emph{nota bene} the $Htt$ Yukawa coupling (of the heaviest
matter particle in the Standard Model) to the Higgs particle.  The
bremsstrahlung of Higgs bosons in the process $e^+e^-\to t\bar t H$
offers a method for measuring this fundamental coupling
directly~\cite{DKLSZ,DR}.  (iii) The Higgs potential which provides the
operational basis for the Higgs mechanism, must finally be
reconstructed.  The strength and the form of the potential
define the Higgs mass and the trilinear and quadrilinear
self-couplings of the Higgs particle.  The prediction for the
trilinear coupling can be tested in Higgs-pair production at the LHC
in the process $pp\to HH$, and at $e^+e^-$ colliders in
Higgs-strahlung $e^+e^-\to ZHH$, for instance, where the Higgs pair is
emitted in the decay of a virtual $H$ boson~\cite{DKMZ}.

\textbf{2.}  Extending the Standard Model to a supersymmetric theory
provides a natural way to keep light fundamental Higgs bosons stable
in the background of large GUT scales.  The search for supersymmetric
particles is therefore a very important endeavor at existing and
future accelerators.  The ultimate discovery limits, in the next two
decades, of squarks and gluinos will be set by the LHC~\cite{Rur}.  TeV
$e^+e^-$ linear colliders will play the same role in the non-colored
sector for charginos/neutralinos and sleptons~\cite{LC}.

Moreover, the high-luminosity version of $e^+e^-$ colliders allows to
carry out very accurate measurements of the sparticle
properties~\cite{LC}.  These measurements can be used to explore
the structure of the  basic supersymmetric theory, in
particular the mechanism responsible for the breaking of
supersymmetry.  Since  SUSY breaking mechanisms are (in general)
generated at scales of the order of the Planck scale, these machines
open windows to physics scenarios in which gravity is an integral part
of the system.  Thus, they prepare the basis for the unification of
the four fundamental forces.

\textbf{3.}  Recent experimental evidence that neutrinos are massive
particles has renewed interest in grand unified theories
such as $SO(10)$, which incorporate right-handed neutrino fields in a
natural way.  Even though the typical scales of  gauge bosons
associated with right-handed symmetries and heavy neutrino
masses may likely be close to the GUT scale, it is nevertheless an
interesting problem to search for new LR degrees of freedom~\cite{CF},
in particular at the LHC which defines the energy frontier of
laboratory accelerators in the near future.

\section{Higgs physics}
\subsection{The Higgs two-photon  channel at the LHC}
The SM Higgs boson in the lower part of the intermediate mass range
$M_H\lesssim 150\;\GeV$, can be searched for at the LHC in the
$H\to\gamma\gamma$ decay mode.  The Higgs bosons are primarily
produced in gluon-gluon fusion.  However, the
signal-to-background ratio improves from typical values $S/B\sim 1/15$
for small-$p_\perp$ Higgs production to $S/B\sim 1/4$ if a cut is
applied to the minimal transverse momentum, balanced by a recoiling
gluon or quark jet~\cite{DIS}:
\begin{equation}
  pp \longrightarrow H(\to\gamma\gamma) + \text{jet}
\end{equation}
Important subprocesses are the parton reactions $g+q/g\to H+q/g$ in
which the Higgs boson is formed in the fusion of the initial-state
gluon and a virtual gluon emitted from the quark/gluon spectator line.

Besides the reducible $\gamma$~backgrounds from decaying hadrons, such
as $\pi^0\to\gamma\gamma$, irreducible backgrounds are generated in QCD
Compton processes like $q+g\to q+\gamma\gamma$; their size can readily
be predicted theoretically~\cite{DIS}.  It is more difficult to control
background processes in which photons are fragments of quarks or
gluons.  These events must be eliminated by vetoing the
accompanying parent jets.  Under realistic experimental conditions the
impurity probabilities are of the order $P[\gamma/q_{\text{veto}}]\sim
2\times 10^{-4}$ and $P[\gamma/g_{\text{veto}}]\sim 0.3\times 10^{-4}$
for quark and gluon parents, respectively.

Selecting large-transverse momentum events allows to search for Higgs
bosons in the $\gamma\gamma$ channel~\cite{CF} also in high-luminosity 
LHC runs
with $\mathcal{L}=100\;\fb^{-1}/\mathrm{year}$ in which many events pile up.  
Exploiting recoiling
high-$p_\perp$ tracks, $z_{\text{vertex}}$ can be nevertheless
 reconstructed very
accurately so that a high $\gamma\gamma$ mass resolution can be
achieved also in this environment: $\sigma_{\gamma\gamma}\approx
690\;\MeV$ in the CMS ECAL, for instance.  With a transverse-momentum
cut of $2\;\GeV$ on the tracks, the average transverse momentum of the
jet formed by the recoiling hadrons is of order $\langle
p_\perp^{\text{jet}}\rangle\approx 40\;\GeV$.

\begin{table}
\caption{Cross section at the LHC, branching ratio
and acceptance for $H \rightarrow \gamma\gamma$ with $M_H = 100$ GeV, 
left column; background cross sections for the same invariant
$\gamma\gamma$ mass, right column.}
{\footnotesize
\begin{displaymath}
\vspace{0.2cm}
\begin{array}{|ll|lrl|} 
\hline 
\multicolumn{2}{|c|}{\mbox{\normalsize Signal $\sigma$}}
&
\multicolumn{3}{|c|}{\mbox{\normalsize Backgrounds $d\sigma/dm_{\gamma\gamma}$}}
\strt\\
\hline
\;\sigma(pp\to H+X) & 56.3\;\pb
&
\;\mbox{quark annihil.} & 92&\,\fb/\GeV\;
\strt\\
\;\text{BR}(H\to\gamma\gamma) & 1.53\times 10^{-3}
&
\;\mbox{gluon fusion} & 167&
\strt\\
\;\mbox{acceptance} & 51.9\,\%
&
\;\mbox{isol.\ bremsstr.} & 120&
\strt\\
\hline
\;{\mbox{\normalsize $\sigma \times$ BR }} & {\mbox {\normalsize 86.2 \; fb}} &
\;{\mbox{\normalsize total }} & {\mbox {\normalsize 379}}
& {\mbox {\, \normalsize fb/GeV}}
\strt\\
\hline
\end{array}
\end{displaymath}
}
\label{tab:2gamma}
\end{table}

\begin{figure}
\begin{center}
\resizebox{8.5cm}{!}{\includegraphics{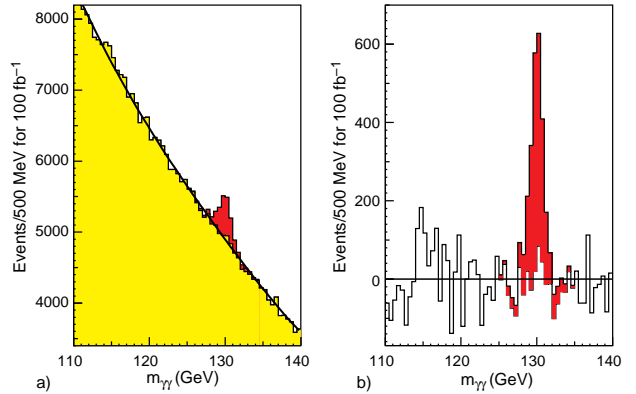}}
\end{center}
\vskip-\baselineskip
\caption[]{The Higgs signal in the
$\gamma\gamma$ channel for $\lumi=100\;\fb^{-1}$ 
before (a) and after (b) background subtraction~\cite{Bar}.}
\label{fig:2gamma}
\end{figure}

The typical size of signal and background cross sections is shown in
Table~\ref{tab:2gamma}, referring to the illustrations in the
Figs.\ref{fig:2gamma}a/b.  The significance $S=N_S/\sqrt{N_B}$ varies
for $\lumi=100\;\fb^{-1}$ from about $8$ to $12$ to $8$ if the Higgs
mass is increased from $100$ to $130$ to $150\;\GeV$, {\it i.e.}
\begin{equation}
  S>8 \quad\text{for}\quad 
  M_H = [100,\, 150\;\GeV]\quad
\end{equation}
Even though theoretical refinements could change the estimates
somewhat, it can nevertheless be concluded that a sufficient
buffer does exist for the discovery of the SM Higgs boson in the
$\gamma\gamma$ resonance channel at the LHC.

\subsection{The parity of Higgs bosons}
The external quantum numbers of the Higgs boson can be studied in
production and decay processes.  For heavy enough Higgs bosons, spin
correlations in Higgs decays to top-quark pairs can be exploited to
measure the parity~\cite{BFH}.  Denoting the couplings
\begin{equation}
  \langle H | t\bar t\rangle
  = (M_t/v)[a + i \gamma_5\tilde a ]
\end{equation}
the coefficient $a\neq 0$ describes the scalar component, the
coefficient $\tilde a\neq 0$ the pseudoscalar component of the state.
If both are non-zero, CP is violated in the interactions.

The couplings determine the spin correlations between the final-state
top and anti-top quark.  Near threshold, the correlations are given by
$\langle s_t s_{\bar t}\rangle = +1/4$ and $-3/4$ for scalar and
pseudoscalar Higgs decay matrix elements, corresponding to spin triplets 
and singlets.  The continuum prediction for the spin correlation in $gg$
collisions at threshold is again $-3/4$.
  The correlation may be analyzed at the
LHC through lepton-angular distributions in semileptonic $t$ and $\bar
t$ decays~\cite{BFH}.  The size of the angular correlations is 
given by $\langle\cos \theta_{\ell^+\ell^-}\rangle$ = 0.396, 
0.383 and 0.402 in the continuum, scalar and
pseudoscalar decays, respectively, for $M_H=400\;\GeV$.
  The statistical error with which
$\langle\cos\theta_{\ell^+\ell^-}\rangle$ can be measured is
estimated to be $\Delta=0.001$ for $\lumi=100\;\fb^{-1}$. 
 Thus, the small differences
between the resonance values and the continuum are significant.

\subsection{The $H tt$ Yukawa coupling}
The masses of the fundamental particles are generated in the Higgs
mechanism by interactions with the non-zero Higgs field in the ground
state.  Within the Standard Model, the Yukawa couplings of the leptons
and quarks with the Higgs particle are therefore uniquely determined
by the particle masses:
\begin{equation}
  g_{Hff}=M_f/v \quad:\quad
  v = (\sqrt2\, G_F)^{-1/2}
\end{equation}
The measurement of these couplings provides a fundamental experimental
test of the Higgs mechanism.  Several methods have
been proposed in the literature to test Yukawa couplings of the Higgs
boson~\cite{LC}.  Since the mass of the top quark is maximal
within the fermion multiplets of the Standard Model, the $Htt$
coupling ranks among the most interesting predictions of the Higgs
mechanism.  This coupling can be tested indirectly by measuring the
$H\gamma\gamma$ and $Hgg$ couplings, which are mediated by virtual
top-quark loops.  If the Higgs boson is very heavy, the decay mode
$H\to t\bar t$ can be exploited.  On the other hand, if the Higgs
boson is light, $M_H\sim 100\ldots 200\;\GeV$, the radiation
of Higgs bosons off top quarks lends itself as a basic mechanism for
measuring the $Htt$ coupling.  The prospects of operating $e^+e^-$
linear colliders at high luminosities\footnote{In TESLA designs,
integrated luminosities of $\lumi=0.3\;\ab^{-1}$ and
$0.5\;\ab^{-1}$ per year are planned at c.m.\ energies
$\sqrt{s}=0.5\;\TeV$ and $0.8\;\TeV$, respectively.}
render the radiation process
\begin{equation}
  e^+e^- \to t\bar t H
\end{equation}
a suitable instrument to carry out this measurement.

\begin{figure}
\begin{center}
\vspace{-2cm}
\resizebox{8cm}{!}{\includegraphics{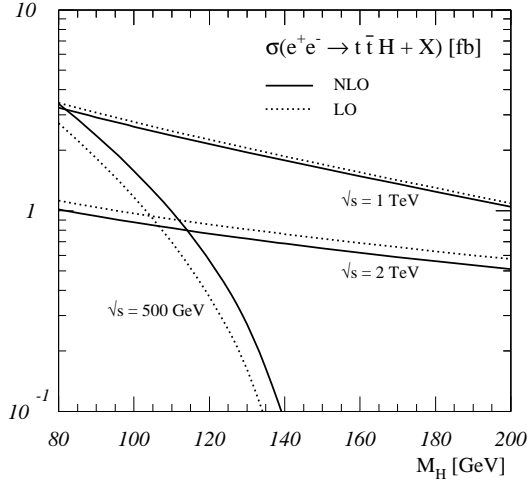}}
\end{center}
\vspace{-1.5\baselineskip}
\caption[]{The cross section of Higgs bremsstrahlung in
$e^+e^- \rightarrow t \bar{t}H$ for three different
collider energies~\cite{DKLSZ}.}
\label{fig:Htt}
\end{figure}

The Higgs boson can be radiated off the final-state top quarks as well
as off the intermediate $Z$~boson; due to the heavy top mass the first
mechanism is by far leading.
%
QCD rescattering corrections~\cite{DKLSZ,DR} increase the cross
section near the threshold while for high energies the positive
contributions from gluon radiation are overwhelmed by negative $Htt$
vertex corrections.  The two limits can be paraphrased by $K$~factors:
\begin{eqnarray}
  K_{\text{thr}}
  &\approx& 1 + \frac{\alpha_s}{\pi}\frac{64}{9}
                \frac{\pi M_t}{\sqrt{(s^{1/2}-M_H)^2-4M^2_t}}
        \\
  K_\infty
  &\approx& 1 - 3\frac{\alpha_s}{\pi}
\end{eqnarray}
The complete cross section including Higgs
emission from the $Z$ line, has been evaluated in Ref.~4.
Characteristic values of the  cross section are 
presented in Fig.\ref{fig:Htt} for three
$e^+e^-$ energy values $\sqrt{s}=500\;\GeV$, $1\;\TeV$ and $2\;\TeV$, as
a function of the Higgs mass.

While the phase-space suppression is gradually lifted with rising
energy, the lower part of the intermediate Higgs mass range is
experimentally accessible already at high-luminosity $e^+e^-$
colliders for $\sqrt{s}=500\;\GeV$; for $\sqrt{s}=1\;\TeV$ the entire
intermediate Higgs mass range can be covered.  For a typical size
$\sigma\sim 1\;\fb$ of the cross section, 
 about $10^3$ events are generated when an integrated
luminosity $\lumi\sim 1\;\ab^{-1}$ is reached within two to three
years of running.  This should provide a sufficiently large sample for
detailed experimental studies of this process.  Since Higgs
radiation from the top quarks is dominant, the sensitivity to the
$Htt$ Yukawa coupling is nearly quadratic,
$\sigma (e^+e^- \to t\bar{t}H) \propto g^2_{Htt}  / 4 \pi$, thus 
being very high.

\subsection{Higgs self-couplings}
The electroweak symmetries are spontaneously broken in the Standard
Model and related theories through a potential in the scalar sector
for which the minimum is realized at a non-vanishing value of the
fields,  
\begin{equation}
  V = \frac{\lambda}{2}\left[ |\varphi|^2 - \frac{v^2}{2}\right]^2
\end{equation}
Expanding the field $\varphi$ about the
ground-state value $\langle\varphi\rangle_0 = v/\sqrt2$, the physical
Higgs mass $M_H=\sqrt{2\lambda}\,v$, and trilinear and quadrilinear
self-couplings of the physical Higgs fields are generated:
\begin{equation}
  V = \frac12 M_H^2 H^2 + \frac12(M_H^2/v)H^3 + \frac18(M_H^2/v^2)H^4
\end{equation}
The fundamental Higgs potential can thus be reconstructed by
measuring the trilinear and quadrilinear self-couplings of the Higgs
boson.

The trilinear self-coupling of the Higgs boson determines the
production of pairs of Higgs particles at $e^+e^-$ colliders and at
the LHC~\cite{DKMZ}.  Several subprocesses are relevant in this context:
\begin{displaymath}
  \begin{array}{l@{\qquad}l@{\quad}l}
  \multicolumn{2}{l}{e^+e^-\to HH+X:} \\[1mm]
  &\mbox{double Higgs-strahlung}
        & e^+e^-\to Z+HH \\
  &\mbox{$WW$ fusion}
        & e^+e^-\to \bar\nu_e\nu_e+HH \\[2mm]
  \multicolumn{2}{l}{pp\to HH+X\;:} \\[1mm]
  &\mbox{double Higgs-strahlung}
        & q\bar q\to W/Z + HH \\
  &\mbox{$WW/ZZ$ fusion}
        & q\bar q\to q\bar q + HH \\
  &\mbox{gluon fusion}
        & gg \to HH
  \end{array}
\end{displaymath}
In all subprocesses, the two Higgs bosons can  either be emitted from
the $W/Z$ lines of the first two subprocesses and the $t$ line of the
third subprocess, or from the splitting of a virtual Higgs boson
$H_{\text{virt}}$ generated, for instance, in Higgs-strahlung
$e^+e^-\to Z+H_{\text{virt}}$ followed by $H_{\text{virt}}\to HH$.

\begin{figure}
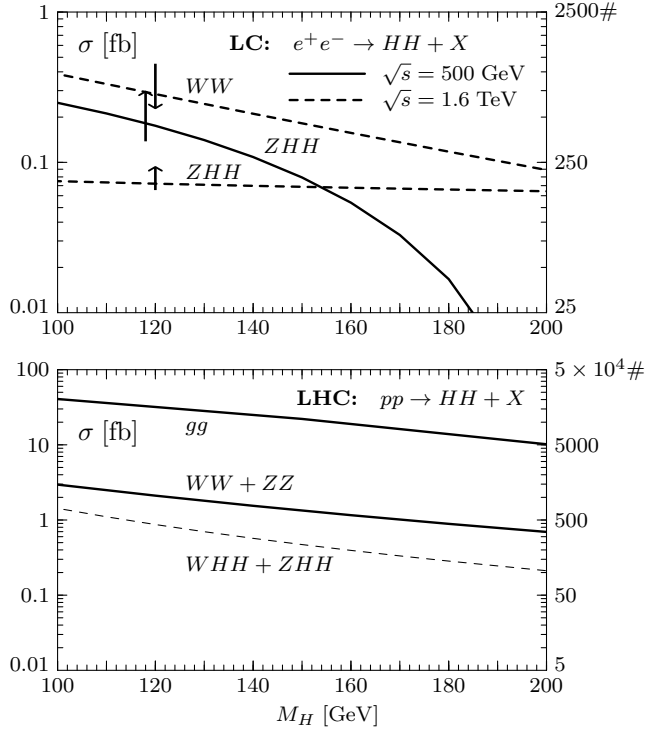

\vspace{2mm}
\begin{center}
\includegraphics{plots_proc.1}\\[7mm]
\includegraphics{plots_proc.2}\\[5mm]
\end{center}
\caption[]{The cross sections for Higgs pair production at 
 $e^+e^-$ colliders and at LHC in various channels. The event rates
on the first and second right scale correspond to integrated luminosities
$\lumi= 2.5\;\ab^{-1}$ and $500\;\fb^{-1}$, respectively.}
\label{fig:HH}
\end{figure}

The size of the cross sections for the different channels is
illustrated in Figs.\ref{fig:HH}a/b.  Since all the processes are of
higher order in the electroweak coupling, the cross sections are
small, posing severe background problems at the LHC, and requiring very 
high luminosities at $e^+e^-$ linear colliders.  The sensitivity to
the trilinear coupling $\lambda_{HHH}$ is indicated by arrows which
point to the variation of the cross sections if the coupling is
modified, \emph{ad hoc}, from $1/2$ to $2\times\lambda_{HHH}$.  The
shifts are distinctly larger than the statistical
fluctuations.

\section{Supersymmetry}

There are strong indications that the Standard Model is embedded in a
supersymmetric theory.  Besides the doubling of the basic Higgs doublet,
SUSY doubles the spectrum of the SM particles in the minimal
supersymmetric extension of the Standard Model (MSSM).  The SM leptons
and quarks are associated with scalar sleptons and squarks; the gluons
with spin-$\frac12$ gluinos; the partners of the electroweak gauge
bosons and Higgs bosons mix to form two charginos and four
neutralinos.  The LHC is the machine proper to search for colored
squarks and gluinos while lepton colliders are suitable machines to
discover the non-strongly interacting particles, charginos/neutralinos
and sleptons.

Apart from variants, essentially two scenarios have been developed to
induce the breaking of supersymmetry: mSUGRA and gauge-mediated
supersymmetry breaking.  Most
phenomenological analyses have been performed so far in the minimal
supergravity model mSUGRA.  In this approach, the breaking is
triggered by gluino condensation in a shadow-world, transferred by
gravitational interactions to the eigen-world.  Soft SUSY breaking
terms are generated this way near Planck scale distances.  Evolving
the universal gaugino and scalar mass parameters from the GUT scale
down to the low electroweak scale, the mass parameters split
eventually, generating the Higgs mechanism when the mass parameter
(squared) in the Higgs sector becomes negative.  This scenario is
described by five parameters, with masses and couplings being
universal at the GUT scale: the scalar
mass $M_0$; the gaugino mass $M_{1/2}$; the trilinear coupling $A_0$;
$\tan\beta$, the ratio of the VEVs in the Higgs sector; and
$\text{sgn}(\mu)$, the sign of the higgsino parameter.  The more
than one hundred phenomenological SUSY parameters at the electroweak scale
can all be expressed in terms of those five fundamental parameters,
leading to  many relations for testing the scheme.

\subsection{Ultimate discovery limits at the LHC}
Sparticles can be searched for at the LHC in a variety of channels.
While the classical signature of squark and gluino production is the
observation of ``$\textit{multi-jets}+E_\perp^{\text{\it miss}}$'', the
search for isolated leptons, together eventually with combinations of
the other signatures, have proven extremely successful,
too~\cite{Rur}:
\begin{equation}
  pp \to \text{sparticles} 
  \to \text{lepton(s)} + (\text{jets}) + E_\perp^{\text{miss}}
\end{equation}
This set of signatures can be applied to the search of
squarks/gluinos, charginos/neutralinos, and sleptons.

\vspace{\baselineskip}
\noindent a) \underline{Squarks and gluinos} \\[.5\baselineskip]
Squark and gluino masses can be expressed in terms of the fundamental
parameters by two simple approximate relations (for moderate values of
$\tan\beta$) in the mSUGRA scenario:
\begin{eqnarray}
  M_{\tilde q} &\approx& \sqrt{M_0^2 + 6M_{1/2}^2} \\
  M_{\tilde g} &\approx& 2.5\,M_{1/2}
\end{eqnarray}
Depending on the relative magnitude of $M_{\tilde q}$ \emph{vs.}
$M_{\tilde g}$, the SUSY particles first decay via mutual cascades,
before leptonic channels become relevant, e.g.
%
\psset{linewidth=.5pt}
\vspace{.5\baselineskip}
\begin{flushleft}
$\qquad
\begin{array}{lll}
pp\;\to\;\rnode{sg1}{\tilde g}\, \rnode{sg2}{\tilde g} / {\tilde g}
{\tilde q} / {\tilde q} {\tilde q} \quad\\
& \rnode{sqq2}{\bar q} + \rnode{sq3}{\tilde q} \quad\\
&& \rnode{q3}{q} + \ell^\pm\nu_\ell + \tilde\chi_1^0 \\
& \rnode{sqq1}{\bar q} + \rnode{sq4}{\tilde q} \\
&& \rnode{q4}{q} + \ell^\pm\nu_\ell + \tilde\chi_1^0
\end{array}
$
\ncangle[angleA=-90,angleB=180,armB=0pt,nodesep=3pt]{->}{sg1}{sqq1} 
\ncangle[angleA=-90,angleB=180,armB=0pt,nodesep=3pt]{->}{sg2}{sqq2}
\ncangle[angleA=-90,angleB=180,armB=0pt,nodesep=3pt]{->}{sq3}{q3} 
\ncangle[angleA=-90,angleB=180,armB=0pt,nodesep=3pt]{->}{sq4}{q4}
\end{flushleft}
\vspace{.5\baselineskip}
\begin{figure}
\begin{center}
\resizebox{6.5cm}{!}{\includegraphics{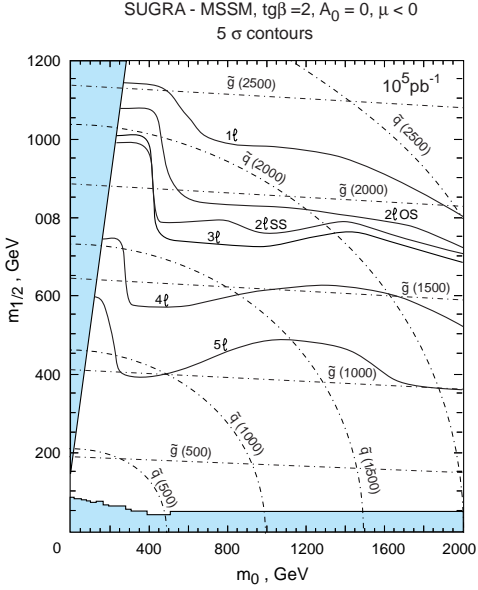}}
\end{center}
\vspace{-1.2\baselineskip}
\caption[]{Contours for the discovery of squarks and
gluinos at the LHC in leptonic channels~\cite{Rur}.}
\label{fig:squark}
\end{figure}
As shown in Fig.\ref{fig:squark}, ultimate discovery limits~\cite{Rur} of
\begin{eqnarray}
  M_{\tilde q} &\lesssim& 2\;\text{to}\;2.5\;\TeV \\
  M_{\tilde g} &\lesssim& 2\;\text{to}\;2.5\;\TeV 
\end{eqnarray}
deep in the TeV range, can be reached for leptonic signatures.  
Similar bounds are found if the parameters $\tan\beta$, $A_0$ and $\mu$ are
altered.

\vspace{\baselineskip}
\noindent b) \underline{Charginos/neutralinos and sleptons}\\[.5\baselineskip]
The mSUGRA mass relations (for the lightest members of the two species)
can be cast in the form:
\vspace{.5\baselineskip}
\begin{flushleft}
$\qquad M_{\tilde\chi_1^0} \approx 0.5\,M_{1/2}
 \qquad\qquad M_{\tilde\chi_1^\pm} \approx M_{1/2}$ \\
$\qquad M_{\tilde\chi_2^0} \approx M_{1/2}$ \\
and\\
$\qquad M_{\tilde\ell_R}^2 
  = M_0^2 + 0.15\,M_{1/2}^2 - s_w^2 M_Z^2 \cos2\beta$ \\
$\qquad M_{\tilde\ell_R}^2
  = M_0^2 + 0.52\,M_{1/2}^2 - \frac12(1-2s_w^2)M_Z^2\cos2\beta$ \\
$\qquad M_{\tilde\nu_L}^2
  = M_0^2 + 0.52\,M_{1/2}^2 + \frac12 M_Z^2 \cos2\beta$
\end{flushleft}
\vspace{.5\baselineskip}
Apart from cascade decays, these particles can be generated through
the Drell-Yan mechanism:
\begin{displaymath}
  pp \to \tilde\chi_i\tilde\chi_j \quad\mbox{and}\quad
         \tilde\ell \bar{\tilde\ell}
\end{displaymath}
in quark-antiquark collisions leading eventually
to clean non--jetty events.  From the abundant leptonic decay
states, bounds~\cite{Rur} of
\vspace{.5\baselineskip}
\begin{flushleft}
$\qquad
\begin{array}{l@{\;:\;}l@{\;\lesssim\;}l}
  \text{charginos/neutralinos} & M_{1/2} & 180\;\GeV \\
  \text{sleptons}              & M_{\tilde\ell}& 350\;\GeV
\end{array}
$
\end{flushleft}
\vspace{.5\baselineskip}
can be set in the high-luminosity runs at LHC.

The mass limits accessible in the Drell-Yan mechanism can be doubled
if suitable cascade decays are realized in the SUSY models.

\subsection{Discovery limits and ultimate precision at LC}
Discovery limits for spin-$\frac12$ $\tilde\chi$ states and 
\mbox{spin-0} $\tilde\ell,\tilde\nu_\ell$ states in $e^+e^-$ 
linear colliders
coincide nearly with the kinematical limits, i.e.
\begin{eqnarray*}
  M_{\tilde\chi_1^\pm} &\lesssim& \sqrt{s}/2 \qquad\text{etc.} \\
  M_{\tilde\ell,\tilde \nu_\ell} &\lesssim& \sqrt{s}/2 \qquad\text{etc.}
\end{eqnarray*}
Thus, in a $2\;\TeV$  collider, chargino/neutralino and
slepton masses up to $\sim 1\;\TeV$ [and in some unpaired
$\tilde\chi_i\tilde\chi_j$ channels even beyond] can be observed.

However, the second target of $e^+e^-$ linear colliders is the
measurement to high precision of the phenomenological SUSY mass
parameters etc.  This is possible in a high-luminosity machine, such
as TESLA, which allows to perform many threshold scans consecutively.
Anticipating~\cite{LC} about $50\;\fb^{-1}$ for the high-precision
mass measurement
at spin-$\frac12$ thresholds which rise steeply $\sim\beta$, and about
$100\;\fb^{-1}$ for spin-0 thresholds which rise more slowly
$\sim\beta^3$, more than a dozen independent channels can be analyzed
if a total luminosity $\lumi \sim 1.5\;\ab^{-1}$ can be provided by the
machine.

In such an experimental program, the masses can be determined at the per-mille level~\cite{LC}:
\begin{eqnarray*}
  M_{\tilde\chi_1^\pm} &=& 138\;\GeV \pm 100\;\MeV \\
  M_{\tilde\mu_R}      &=& 132\;\GeV \pm 300\;\MeV
  \qquad\text{etc.}
\end{eqnarray*}
Based on these measurements, the fundamental mSUGRA parameters can be
extracted with very high accuracy~\cite{Bla}; for example,
\vspace{.5\baselineskip}
\begin{flushleft}
$\qquad
\begin{array}{l@{\;=\;}l@{\quad}l@{\;=\;}l}
  \tan\beta & 3\pm 0.01    & M_0     & 100\;\GeV \pm 120\;\MeV \\
  A_0       & 0\pm 5\;\GeV & M_{1/2} & 200\;\GeV \pm 130\;\MeV
\end{array}
$
\end{flushleft}
\vspace{.5\baselineskip}
These high-precision measurements at the per-mille level will allow us
to test stringently scenarios of supersymmetry breaking.  Since these
mechanisms are generated at the GUT/Planck scale, eventually involving
gravity, high-luminosity $e^+e^-$ linear colliders appear
indispensable facilities for opening windows to physics scales where gravity
plays an integral role, thus providing a bridge to the unification of
all four forces.

\section{Left-right symmetric gauge theories}
Left-right symmetric gauge theories connect the right-chiral leptons,
charged and neutral, and the right-chiral quarks by the absorption or
emission of $W_R^\pm$ gauge bosons.  Moreover, neutral gauge bosons
$Z'$ exist in these scenarios, heavier than the $W_R^\pm$ by the ratio
$(2\cos^2\theta_w/\cos2\theta_w)^{1/2} \approx 1.7$.  The left and
right degrees of freedom mix in general to form mass eigenstates
$W,W'$ and $Z,Z'$; however, these mixing effects will in general be
neglected. 

To generate a spectrum of very light and very heavy neutrinos, a
see-saw mechanism may be operative which is driven by  large Majorana
masses associated with the right-handed neutrinos.  The light as well as
the heavy mass eigenstates are Majorana neutrinos $\nu$ and $N$, marked
by a family index each.

The gauge bosons $W_R^\pm$ and $Z'$ can be produced in the Drell-Yan
mechanism:
\begin{equation}
  pp \to \mbox{$W_R$ and $Z'$}
\end{equation}
The $W_R$ bosons are produced in collisions of right-handed quarks and
left-handed antiquarks.  Similarly $Z'$, yet also in quark/antiquark
beams with reverse handedness.  The particles decay into
right/left-handed quarks/antiquarks following the same rule, yet also
to charged leptons and heavy neutrinos, if kinematically possible:
\vspace{.5\baselineskip}
\begin{flushleft}
$\qquad
\begin{array}{l@{\;\to\;}l@{\quad\text{and}\quad}l}
  W_R^- & q\bar q' & N_{\ell}\ell^- \\
  Z'    & q\bar q  & \ell^+\ell^- \oplus N_{\ell}N_{\ell}
\end{array}
$
\end{flushleft}
\vspace{.5\baselineskip}
The quark decays dominate strongly.  The Majorana neutrinos $N$ decay,
if mixing effects are neglected, into charged leptons
of either sign with
equal probability,
\begin{equation}
  N_{\ell} \to \ell^\pm + jj
\end{equation}
where a right-handed quark current is coupled to the right-handed
lepton current by virtual $W_R$ exchange.

\begin{table}
\caption[]{Discovery limits of gauge bosons and heavy Majorana 
neutrinos in LR symmetric  theories at the LHC~\cite{CF}.}
{\begin{displaymath}
\vspace{0.2cm}
\begin{array}{|l|cc|} 
\hline 
\;pp \to & \;M(W_R/Z')\; & M(N) \strt\\
\hline
\;W_R \to \ell N_\ell\; & 6.4\;\TeV & \;3.3\;\TeV\; \strt\\
\;Z' \to N_\ell N_\ell\; & 4.5\;\TeV & \;1.7\;\TeV\; \strt\\ 
\hline
\end{array}
\end{displaymath}
}
\label{tab:LR}
\end{table}

The signatures of the events~\cite{CF} are spectacular isolated leptons
$\ell^\pm$ plus two jets, clustering at the $N_{\ell} = (\ell jj)$ mass,
 and a second
charged lepton, clustering together with $N_{\ell}$ at the
$W_R=(\ell\ell jj)$ mass.  Similarly the $Z'$ chains.  As expected,
the $W_R$ channel provides the highest discovery limit~\cite{CF}, shown in
Table~\ref{tab:LR}.

Thus, left-right symmetric phenomena can be probed at the LHC in the
multi-TeV mass range, in the gauge-boson as well as the neutrino
sector.

\section*{Acknowledgements}
We are grateful to R.\ McPherson and H.\ Weerts for the invitation
to this talk. Thanks should go to 
D.\ Barney, W.\ Bernreuther and L.\ Rurua for
providing us with unpublished material.  We also thank our
collaborators G.\ Blair, A.\ Djouadi and M.\ Muhlleitner for allowing us to
include unpublished results. Help in preparing the manuscript is 
acknowledged to  S.\ G\"unther and T.\ Plehn.


\end{document}